\documentclass[aps,twocolumn,showpacs,preprintnumbers,amsmath,amssymb]{revtex4}

\usepackage{graphicx}
\usepackage{dcolumn}
\usepackage{bm}
\usepackage{hyperref}

\newcommand{\beq}{\begin{equation}}
\newcommand{\eeq}{\end{equation}}
\newcommand{\beqa}{\begin{eqnarray}}
\newcommand{\eeqa}{\end{eqnarray}}
\newcommand{\beqar}{\begin{eqnarray*}}
\newcommand{\eeqar}{\end{eqnarray*}}
\newcommand{\bra}[1]{\mbox{$\left\langle{#1}\right|$}}
\newcommand{\ket}[1]{\mbox{$\left|{#1}\right\rangle$}}
\newcommand{\diracsp}[2]{\mbox{$\langle{#1}|{#2}\rangle$}}

\newcounter{saveeqn}
\newcommand{\alpheqn}{\setcounter{saveeqn}{\value{equation}}%
\stepcounter{saveeqn}\setcounter{equation}{0}%
\renewcommand{\theequation}{\mbox{\arabic{saveeqn}\alph{equation}}}}
\newcommand{\reseteqn}{\setcounter{equation}{\value{saveeqn}}%
\renewcommand{\theequation}{\arabic{equation}}}
\def\beql{\alpheqn \beqa}
\def\eeql{\eeqa \reseteqn}

\begin{document}

\title{Generalized GHZ-class and W-class concurrence and entanglement vectors \\
of the multipartite systems consisting of qubits}
\author{An Min WANG$^{1,2}$}


\altaffiliation{It is founded by the National Fundamental Research Program of China
with No. 2001CB309310, partially supported by the National Natural Science Foundation of China
under Grant No. 60173047  and the Natural Science Foundation of Anhui Province. \\
}

\affiliation{$^{1}$ Laboratory of Quantum Communication and
Quantum Computing and Institute for Theoretical Physics}
\affiliation{$^{2}$Department of Modern Physics, University of
Science and Technology of China, Hefei 230026, People's Republic
of China}



\begin{abstract}
We propose two classes of the generalized concurrence vectors of
the multipartite systems consisting of qubits. Making use of them,
we are able to, respectively, describe and quantify GHZ-class and
W-class entanglement both in total and between arbitrary two
partite in the multipartite system consisting of qubits. In the
case of pure state of three qubits that one partite is separable,
it is shown to exactly back to the usual Wootters' concurrence
after introduce a whole concurrence vector. In principle, our
method is applicable to any $N$-partite systems consisting of $N$
qubits.
\end{abstract}


\maketitle

Quantum entanglement lies at the heart of quantum mechanics and is
viewed as a useful resource in quantum information. At present,
there are three most promising ideas for quantifying entanglement.
They are the entanglement of formation \cite{Bennett1,Bennett2},
the relative entropy of entanglement \cite{Vedral1} and the
entanglement of distillation \cite{Bennett1, Rains} for bipartite
systems. In particular, {\it via} Wootters' development
\cite{Wootters}, the entanglement of formation is written as a
binary entropy function of the density matrix or concurrence for a
pair of qubits.

Since Bennett {\it et. al}'s work \cite{Bennett3}, the measures of
entanglement in multipartite systems have attracted a lot of
attention and achieved some developments. One of them is that a
kind of form of concurrence of three qubits was suggested
\cite{3qC}.

Recent years, based on the above ideas of the entanglement
measures, we have tried to put forward to the generalized
entanglement of formation (GEF)\cite{MyEFmp} and modified relative
entropy of entanglement (MRE) in multipartite systems
\cite{MyMRE}. Moreover, we also study the obvious expression of
concurrence \cite{MyOEC} and the general solution of a $4\times 4$
Hermit and positive matrix \cite{MyGS2qDM} in order to seek for a
general expression of concurrence of multipartite systems
consisting of qubits. However, it is still difficult to find out
such a form of concurrence of many qubits since involving amount
of calculations.

Usually, for simplicity, the entanglement of multipartite systems
is thought of a single scalar. However, because that three qubits
can be entangled in two inequivalent ways and one class state of
them can not be obtained from another class state $via.$ any local
operation and classical communication (LOCC)\cite{TwoClassE}, we
can think that the entanglement measure contains or divides two
parts: GHZ-class and W-class ones. In addition, the entanglement
between arbitrary two partite also need be known in practise.
Thus, in our view, it is better to use the vectorial and
class-related measures to quantify the entanglement of
multipartite systems. Therefore, in this letter, we propose two
classes of generalized concurrence vectors of the multipartite
systems consisting of qubits, which are functions of the density
of matrix. Making use of them, we are able to, respectively,
describe and qualify GHZ-class and W-class entanglement both in
total and between arbitrary two partite in the multipartite
systems consisting of qubits. In the case of pure state of three
qubits that one partite is separable, it is shown to exactly back
to the usual Wootters' concurrence after introduce a whole
concurrence vector. In princple, our method is applicable to any
$N$-partite system consisting of $N$ qubits.

Let us start with the concurrence of two qubits
\cite{Bennett2,Wootters}. In terms of so-called magic basis, one
can rewrite the entanglement of formation for a pure state
$\ket{\psi}=a\ket{00}+b\ket{01}+c\ket{10}+d\ket{11}$ as a binary
entropy function \beql \label{bef}
H(x)&=&-x\log x-(1-x)\log(1-x), \\
x&=&\frac{1+\sqrt{1-C^2}}{2} \eeql where $C$ is called as
``concurrence", and \beq C=2|ad-bc| \eeq

The idea of Wootters' work \cite{Wootters} is just to take such a
binary entropy function as a measure of entanglement for mixed
state and obtain a method to calculate correct concurrence so that
this binary entropy function dependent on $C$  is  not larger than
$\sum_ip^i E_{EF}(\rho_i^{\rm P})$ for any pure state
decompositions $\rho^{\rm M}=\sum_i p^i\rho_i^{\rm P}$.  Wootters
gave out, for the mixed state $\rho$ of two qubits, the
concurrence is \beq
C=\max(\lambda_1-\lambda_2-\lambda_3-\lambda_4,0) \eeq where the
$\lambda_i$, in decreasing order, are the square roots of the
eigenvalues of the generation matrix$
\rho^{1/2}\tilde{\rho}\rho^{1/2}=\rho^{1/2}(\sigma_2\otimes
\sigma_2){\rho}^*(\sigma_2\otimes\sigma_2)\rho^{1/2}$, and
$\rho^*$ denotes the complex conjugation of $\rho$ in the
computation basis $\{\ket{00},\ket{01},\ket{10},\ket{11}\}$.
Alternatively, one can say that the $\lambda_{i}$ are the square
roots of the eigenvalues of $\rho\tilde{\rho}$.

Wootters' formula of concurrence makes use of what can be called
``spin flip" transformation. For a general state described by a
density matrix $\rho$ in the systems of two qubits, this
spin-flipped state is defined by \beq
\tilde{\rho}_{A_1\!A_2}=(\sigma_2\otimes\sigma_2)\rho^*_{A_1\!A_2}(\sigma_2\otimes\sigma_2)
\eeq One has known that this way can not be directly extended to
the multipartite systems because the trace of $\rho\tilde{\rho}$
is zero for a pure state of three qubits.

In order to overcome this difficulty, we note that
$\sigma_2\otimes\sigma_2$ can be written as a linear combination
of Bell's states with the weight $\pm 1$, that is  \beq
\sigma_2\otimes\sigma_2=\sum_{i=1}^4 c_i \ket{B_i}\bra{B_i} \eeq
where $c_1=c_4=-1, c_2=c_3=1$, and four Bell's states $\ket{B_i}$
are denoted by, respectively,
$\ket{B_1}=(\ket{00}+\ket{11})/\sqrt{2}$,
$\ket{B_2}=(\ket{00}-\ket{11})/\sqrt{2}$,
$\ket{B_3}=(\ket{01}+\ket{10})/\sqrt{2}$,
$\ket{B_4}=(\ket{01}-\ket{10})/\sqrt{2}$. Therefore, we can
rewrite the ``spin flipped" (tilde) density matrix as \beq
\tilde{\rho}_{A_1\!A_2}=\sum_{i,j=1}^4 c_i
c_j\bra{B_j}\rho_{A_1\!A_2}\ket{B_i}\ket{B_i}\bra{B_j} \eeq It
implies that the generation matrix of concurrence
$\rho^{1/2}\tilde{\rho}\rho^{1/2}$ has the following the important
physical features: (1) Its matrix elements are the linear
combinations of the matrix elements of density matrix on cat state
basis with the weight $\pm 1$; (2) It is just a cat state when the
$\rho$ is a cat state; (3) It is a zero matrix when the $\rho$ is
a (fully) separable pure state.

Our physical idea (or assumption) is the above three features are
kept in the case of multipartite systems consisting of qubits.
However, from the above statement, such a concurrence must be a
class of concurrence that only can quantify GHZ-class entanglement
of the entangled states, which can be seen more clearly later.

In order to obtain the concurrence of multipartite of systems,
let's first denote the GHZ's states for $N$-qubits

\beq \!\!\!\!\ket{g^{\rm N}_i}=\frac{1}{\sqrt{2}}\left(\ket{s^{\rm
N}_{\left[(i+1)/2\right]}}+(-1)^{i-1}\ket{s^{\rm
N}_{2^N-1+\left[(i+1)/2\right]}}\right) \eeq where $[\ ]$ means to
take the integer part and all of $\ket{s^{\rm N}_i}$ are the
familiar spin basis (computation basis).

Consider the case of three qubits, its ``tilde" state should be
written as \beq \label{tdmghz}\tilde{\rho}^{\rm
ghz}_{A_1\!A_2\!A_3}=\sum_{i,j=1}^8 c_i
c_j\bra{g^3_j}\rho_{A_1\!A_2\!A_3}\ket{g^3_i}\ket{g^3_i}\bra{g^3_j}
\eeq In particular, for a pure state $\rho=\ket{\psi}\bra{\psi}$,
we have \beq \label{tildeghz}\tilde{\rho}^{\rm
ghz}_{A_1\!A_2\!A_3}=\ket{\Psi}\bra{\Psi},\quad
\ket{\Psi}=\sum_{i=1}^8 c_i\diracsp{\psi}{g^3_i}\ket{g^3_i}\eeq

Under the requirement keeping the invariance of GHZ's states, we
have $c_i^2=1$, that is, $c_i$ is equal to $1$ or $-1$. Set
$\rho=\ket{000}\bra{000}$, from $\tilde{\rho}\rho=0$ it follows
that $c_1=-c_2$. This is because that $ \ket{\Psi}$ becomes
$(c_1\ket{g^3_1}+c2\ket{g^3_2})/\sqrt{2}$ and $
\diracsp{\Psi}{000}=0$ result in $c1+c2=0$. Likewise, we obtain
\beq\label{crelation} c_{2i-1}=-c_{2i}\quad (i=1,2,3,4)\eeq

Furthermore, when $\rho$ is a fully separable pure state, without
loss of generality, we write it by \beq
\ket{\psi_s}\!\!\!=\!\!(a_1\ket{0}+b_1\ket{1})\otimes(a_2\ket{0}
+b_2\ket{1})\otimes(a_3\ket{0}+b_3\ket{1}) \eeq Thus, from
Eq.(\ref{tildeghz}) it follows that \beq
\diracsp{\Psi}{\psi_s}=2(c_1+c_3+c_5+c_7)a_1a_2a_3b_1b_2b_3 \eeq
Based on our precondition, for any fully separable pure state,
$\diracsp{\Psi}{\psi_s}=0$ , it follows that, \beq
c_1+c_3+c_5+c_7=0 \eeq It corresponds to three choices: (1)
$c_1=c_3=-1$, $c_5=c_7=1$; (2)$c_1=c_5=-1$, $c_3=c_7=1$;
(3)$c_1=c_7=-1$, $c_3=c_5=1$. The others are given by
eq.(\ref{crelation}).

{\em It must be emphasized that $\tilde{\rho}$ has to be get by a
local transformation from $\rho$. Otherwise, it can change the
quantity of entanglement of the studied state in general}.

Actually, we have found out three local transformations which are
the direct products of Pauli matrices \beql \tilde{\rho}^{\rm
ghz}(1,2)=(\sigma_2\otimes\sigma_2\otimes
\sigma_1)\rho^* (\sigma_2\otimes\sigma_2\otimes\sigma_1)\\
\tilde{\rho}^{\rm
ghz}(1,3)=(\sigma_2\otimes\sigma_1\otimes\sigma_2)\rho^*
(\sigma_2\otimes\sigma_1\otimes\sigma_2)\\ \tilde{\rho}^{\rm
ghz}(2,3)=(\sigma_1\otimes\sigma_2\otimes\sigma_2)\rho^*
(\sigma_1\otimes\sigma_2\otimes\sigma_2)\eeql Because these
product of Pauli matrices have forms \beq
\sum_{k=1}^8c_k(i,j)\ket{g^3_k}\bra{g^3_k} \quad (i<j,\
i,j=1,2,3)\eeq and their coefficients $c_k(1,2), c_k(1,3)$ and
$c_k(2,3)$ respectively correspond to the three choices (1), (2)
and (3).

Therefore, we can define the generation matrices of GHZ-class
concurrences by \beq \!\!\!M_C^{\rm
ghz}(i,j)=\rho^{1/2}\tilde{\rho}^{\rm ghz}(i,j)\rho^{1/2}\eeq

First, consider a pure state $\rho=\ket{\psi}\bra{\psi}$. Because
$\rho^2=\rho$, then $\rho^{1/2}=\rho$. The generation matrix of
concurrence becomes $M_C=|\diracsp{\Psi}{\psi}|^2\rho$. It means
that {\it for a pure state $\ket{\psi}$, its three {\rm GHZ}-class
concurrences are given by} \beq \!\!\!C^{\rm
ghz}_{ij}=|\diracsp{\Psi_{ij}}{\psi}|,\quad
\ket{\Psi_{ij}}=\sum_{k=1}^8
c_k(i,j)\diracsp{\psi}{g^3_k}\ket{g^3_k}\eeq

For the case of mixed states, we suggest according to Wootters's
form, without a strict proof, in multipartite systems, and then
defining the concurrences \beq \!\!\!C^{\rm
ghz}_{ij}=\max\!\left(2\lambda_1^{\rm
ghz}(i,j)-\sum_{k=1}^8\lambda_k^{\rm ghz}(i,j),0\right) \eeq where
the $\lambda_k^{\rm ghz}(i,j)$ $(i<j,\ i,j=1,2,3)$, in decreasing
order, are the square roots of the eigenvalues corresponding to
the generation matrices $M_C^{\rm ghz}(1,2),M_C^{\rm ghz}(1,3)$
and $M_C^{\rm ghz}(2,3)$.

We think that the above three concurrences form a so-called
GHZ-class ``concurrence vector" of the entangled states. Making
use of $C^{\rm ghz}_{ij},\ (i<j, i,j=1,2,3)$, we can rightly
describe and qualify GHZ-class entanglement of the entangled
states between the partite $i$ and the partite $j$ {\it via.}
defining a GHZ-class measure vector of entanglement, whose
components are \beq E^{\rm
ghz}_{ij}=H\left[\frac{1}{2}\left(1+\sqrt{1-C^{\rm
ghz}_{ij}{}^2}\right)\right]\eeq where $H(x)$ is a binary entropy
function as Eq.(\ref{bef}). The total GHZ-class entanglement
measure, as a scalar, can be defined as the norm of this
entanglement vector \beq E^{\rm ghz}_{T}=\sqrt{E^{\rm
ghz}_{12}{}^2+E^{\rm ghz}_{13}{}^2+E^{\rm ghz}_{23}{}^2} \eeq

Without loss of generality, we denote a pure state by \beq
\label{3qps}\ket{\psi}=\sum_{i=1}^8 x_i\ket{s^3_i}\eeq It is easy
to obtain that \beql C^{\rm ghz}_{12}&=&2|
x_4 x_5 + x_3 x_6 - x_2 x_7 - x_1 x_8|\\
C^{\rm ghz}_{13}&=&2| x_4 x_5 - x_3 x_6 + x_2 x_7 - x_1
x_8|\\C^{\rm ghz}_{23}&=&2| x_4 x_5 - x_3 x_6 - x_2 x_7 + x_1 x_8|
\eeql Obviously, for the fully separable pure states, they are
zero. For the general GHZ's states including the non maximum ones
(for example, $\alpha\ket{000}+\beta\ket{111}$), their components
of concurrence vector are all equal to $2|\alpha\beta|$. However,
for a general W-state or a general anti W-state: \beqa\label{ws}
\ket{W}&=&\alpha\ket{001}+\beta\ket{010}+\gamma\ket{100}\\
\label{antiws}
\ket{\overline{W}}&=&\alpha\ket{011}+\beta\ket{101}+\gamma\ket{110}\eeqa
their above components of concurrence vector are equal to zero. It
implies that the above concurrence vector can not qualify the
entanglement of W-class states rightly. It is not surprised since
our above definition in Eq.(\ref{tdmghz}).

In terms of similar physical idea and calculation method, we are
able to define the W-class concurrence vector. First, we write
down \beql \tilde{\rho}^{\rm
W}(1,2)&=&(\sigma_2\otimes\sigma_2\otimes
\sigma_0)\rho^* (\sigma_2\otimes\sigma_2\otimes\sigma_0)\\
\tilde{\rho}^{\rm
W}(1,3)&=&(\sigma_2\otimes\sigma_0\otimes\sigma_2)\rho^*
(\sigma_2\otimes\sigma_0\otimes\sigma_2)\\
 \tilde{\rho}^{\rm
W}(2,3)&=&(\sigma_0\otimes\sigma_2\otimes\sigma_2)\rho^*
(\sigma_1\otimes\sigma_2\otimes\sigma_2)\eeql where we have used
the fact that these local transformations have the forms \beq
\sum_{k=1}^8 d_k(i,j)\ket{B_k(i,j)}\bra{B_k(i,j)}\eeq where
$\ket{B_k(i,j)}$ $(i<j,\ i,j=1,2,3)$ are that Bell's states of
$(i,j)$-partite product a spin basis of the third sparable
partite, for example, \beql
\ket{B_1(1,2)}&=&(\ket{s^3_1}+\ket{s^3_7})/\sqrt{2}\\
\ket{B_2(1,2)}&=&(\ket{s^3_1}-\ket{s^3_7})/\sqrt{2}\\
\ket{B_3(1,2)}&=&(\ket{s^3_3}+\ket{s^3_5})/\sqrt{2}\\
\ket{B_4(1,2)}&=&(\ket{s^3_3}-\ket{s^3_5})/\sqrt{2}\\
\ket{B_5(1,2)}&=&(\ket{s^3_2}+\ket{s^3_8})/\sqrt{2}\\
\ket{B_6(1,2)}&=&(\ket{s^3_2}-\ket{s^3_8})/\sqrt{2}\\
\ket{B_7(1,2)}&=&(\ket{s^3_4}+\ket{s^3_6})/\sqrt{2}\\
\ket{B_8(1,2)}&=&(\ket{s^3_4}-\ket{s^3_6})/\sqrt{2} \eeql and the
weights are \beql
\!\!\!d_1(i,j)=-d_3(i,j)\!=\!-d_5(i,j)=d_7(i,j)=-1\\
d_{2k-1}(i,j)\!=\!-d_{2k}(i,j) \quad (i<j,\ i,j=1,2,3)\eeql
Therefore, we can define the generation matrices of W-class
concurrence vector by \beq M_C^{\rm
W}(i,j)=\rho^{1/2}\tilde{\rho}^{\rm W}(i,j)\rho^{1/2}\eeq It is
easy to obtain that {\it for a pure state $\ket{\psi}$, its
W-class concurrence is given by} \beql C^{\rm
W}_{ij}&=&|\diracsp{\Psi_{ij}}{\psi}|\\
 \ket{\Psi_{ij}}&=&\sum_{k=1}^8
d_k(i,j)\diracsp{\psi}{B_k(i,j)}\ket{B_k(i,j)})\eeql While W-class
concurrence vector for the mixed states, again according to
Wootters' form without proof, can be suggested as \beq C^{\rm
W}_{ij}=\max\left(2\lambda_1^{\rm
W}(i,j)-\sum_{k=1}^8\lambda_k^{\rm W}(i,j),0\right) \eeq where the
$\lambda_k^{\rm W}(i,j)$ $(i<j,\ i,j=1,2,3)$, in decreasing order,
are the square roots of the eigenvalues corresponding to the
generation matrices $M_C^{\rm W}(1,2),M_C^{\rm W}(1,3)$ and
$M_C^{\rm W}(2,3)$. Then a W-class measure vector of entanglement
and its norm are just \beqa E^{\rm
W}_{ij}&=&H\left[\frac{1}{2}\left(1+\sqrt{1-C^{\rm
W}_{ij}{}^2}\right)\right]\\ E^{\rm W}_{T}&=&\sqrt{E^{\rm
W}_{12}{}^2+E^{\rm W}_{13}{}^2+E^{\rm W}_{23}{}^2} \eeqa They can
be used to describe and qualify W-class entanglement of the
entangled states both in total and between arbitrary two partite.

It is easy to get that for a pure state defined by Eq.(\ref{3qps})
\beql
C^{\rm W}_{12}&=&2|x_3x_5 + x_4 x_6 - x_1 x_7 - x_2 x_8|\\
C^{\rm W}_{13}&=&2|x_2x_5 - x_1x_6 + x_4x_7 - x_3 x_8|\\
C^{\rm W}_{23}&=&2|x_2x_3 - x_1 x_4 + x_6 x_7 - x_5 x_8| \eeql For
the fully separable pure states, they are zero. For a general
W-state and anti W-state, we have \beql C^{\rm
W}_{12}=2|\beta\gamma|,&\quad& C^{\rm
W}_{13}=2|\alpha\gamma|,\quad C^{\rm W}_{23}=2|\alpha\beta|\\
C^{\rm \overline{W}}_{12}=2|\alpha\beta|,&\quad& C^{\rm
\overline{W}}_{13}=2|\alpha\gamma|,\quad
 C^{\rm\overline{W}}_{23}=2|\beta\gamma| \eeql
They rightly give out the pair of entanglement of the general
W-states and/or the general anti W-states \cite{TwoClassE}. For
the general GHZ's states including the non maximum ones (for
example, $\alpha\ket{000}+\beta\ket{111}$), their components of
W-class concurrence vector are all equal to 0. Again, note that
the fact that GHZ-class concurrence vectors of the general
W-states or the general anti W-states are equal to zero, we think
that these results imply that our classification to be useful and
correct.

{\em It must point out that both $C^{\rm ghz}$ and $C^{\rm W}$ do
not exactly describe the partially separable states in general}.
As a example, considering the state
$(y_1\ket{00}+y_2\ket{01}+y_3\ket{10}+y_4\ket{11})\otimes(a_3\ket{0}+b_3\ket{1})$.
In terms of our concurrences, its non-zero components of
concurrence vectors are \beql
C^{\rm{ghz}}_{12}&=&2|y_1y_4-y_2y_3|(2|a_3b_3|)\\
C^{\rm{W}}_{12}&=&2|y_1y_4-y_2y_3|(|a_3^2+b_3^2|)\eeql the others
components are zero as expected. In order to exactly back to the
usual case of two qubits at least for a pure state, we introduce
the whole concurrence vector\beq \label{wconcurrence}
{\cal{C}}_{ij}=\frac{C_{ij}^{\rm ghz}+C_{ij}^{\rm W}}{\sum_\alpha
\lambda_\alpha
(|a_k^2(\alpha)+b_k^2(\alpha)|+2|a_k(\alpha)b_k(\alpha)|)}\eeq
where $i\neq j\neq k,\ i<j$, and $\lambda_\alpha$ and
$a_k(\alpha)\ket{0}+b_k(\alpha)\ket{1}$ are respectively the
eigenvalues and corresponding eigenvectors of the reduced density
matrix of $k$ partite. {\it Actually, in the definitions of
$C_{ij}^{\rm ghz}$ and $C_{ij}^{\rm W}$, the above denominator in
eq.(\ref{wconcurrence}) can be added either}. In our above
example, it is $|a_3^2+b_3^2|+2|a_3b_3|$, and thus
${\cal{C}}_{12}=2|y_1y_4-y_2y_3|$. It is just the same as
Wootters' concurrence \cite{Wootters}. Therefore, we can think
that the whole entanglement vector of three qubits can be defined
by \beq
E_{ij}=H\left[\frac{1}{2}\left(1+\sqrt{1-{\cal{C}}_{ij}{}^2}\right)\right]
\eeq and its norm, as a measure of whole entanglement, is just
\beq \label{wentanglement}
E=\sqrt{E_{12}{}^2+E_{13}{}^2+E_{23}{}^2} \eeq

For four qubits, we can, in similar way, define \beql
\!\!\!\tilde{\rho}^{\rm
ghz}(1,2)\!\!=\!\!(\sigma_2\!\otimes\!\sigma_2\!\otimes\!
\sigma_1\!\otimes\!\sigma_1)\rho^* (\sigma_2\!\otimes\!\sigma_2\!\otimes\!\sigma_1\!\otimes\!\sigma_1)\\
\!\!\!\tilde{\rho}^{\rm
ghz}(1,3)\!\!=\!\!(\sigma_2\!\otimes\!\sigma_1\!\otimes\!\sigma_2\!\otimes\!\sigma_1)\rho^*
(\sigma_2\!\otimes\sigma_1\!\otimes\!\sigma_2\!\otimes\!\sigma_1)\\
\!\!\!\tilde{\rho}^{\rm
ghz}(1,4)\!\!=\!\!(\sigma_2\!\otimes\!\sigma_1\!\otimes\!\sigma_1\!\otimes\!\sigma_2)\rho^*
(\sigma_2\!\otimes\!\sigma_1\!\otimes\!\sigma_1\!\otimes\!\sigma_2)\\
\!\!\!\tilde{\rho}^{\rm
ghz}(2,3)\!\!=\!\!(\sigma_1\!\otimes\!\sigma_2\!\otimes\!\sigma_2\!\otimes\!\sigma_1)\rho^*
(\sigma_1\!\otimes\!\sigma_2\!\otimes\!\sigma_2\!\otimes\!\sigma_1)\\
\!\!\!\tilde{\rho}^{\rm
ghz}(2,4)\!\!=\!\!(\sigma_1\!\otimes\!\sigma_2\!\otimes\!\sigma_1\!\otimes\!\sigma_2)\rho^*
(\sigma_1\!\otimes\!\sigma_2\!\otimes\!\sigma_1\!\otimes\!\sigma_2)\eeql
\beql \!\!\!\tilde{\rho}^{\rm
W}(1,2)\!\!=\!\!(\sigma_2\!\otimes\!\sigma_2\!\otimes\!
\sigma_0\!\otimes\!\sigma_0)\rho^* (\sigma_2\!\otimes\!\sigma_2\!\otimes\!\sigma_0\!\otimes\!\sigma_0)\\
\!\!\!\tilde{\rho}^{\rm
W}(1,3)\!\!=\!\!(\sigma_2\!\otimes\!\sigma_0\!\otimes\!\sigma_2\!\otimes\!\sigma_0)\rho^*
(\sigma_2\!\otimes\sigma_0\!\otimes\!\sigma_2\!\otimes\!\sigma_0)\\
\!\!\!\tilde{\rho}^{\rm
W}(1,4)\!\!=\!\!(\sigma_2\!\otimes\!\sigma_0\!\otimes\!\sigma_0\!\otimes\!\sigma_2)\rho^*
(\sigma_2\!\otimes\!\sigma_0\!\otimes\!\sigma_0\!\otimes\!\sigma_2)\\
\!\!\!\tilde{\rho}^{\rm
W}(2,3)\!\!=\!\!(\sigma_0\!\otimes\!\sigma_2\!\otimes\!\sigma_2\!\otimes\!\sigma_0)\rho^*
(\sigma_0\!\otimes\!\sigma_2\!\otimes\!\sigma_2\!\otimes\!\sigma_0)\\
\!\!\!\tilde{\rho}^{\rm
W}(2,4)\!\!=\!\!(\sigma_0\!\otimes\!\sigma_2\!\otimes\!\sigma_0\!\otimes\!\sigma_2)\rho^*
(\sigma_0\!\otimes\!\sigma_2\!\otimes\!\sigma_0\!\otimes\!\sigma_2)\eeql

Further, in terms of the above methods and procedure, we can
define the two-classes of concurrence vectors and then the two
classes of entanglement vectors which respectively quantify
GHZ-class and W-class entanglement of the entangled states.
Obviously, we can extend, in princple, our method to any
$N$-partite system consisting of $N$ qubits.

In the end, we would like to point out that our physical idea is
reasonable and our conclusions are clear at least for the pure
states. The whole entanglement also can be thought of
eq.(\ref{wentanglement}) when the classification is not important.
In the case of mixed states, a strict proof about the definitions
of two classes of concurrence vectors is expected. It seems to be
very difficult since so far ones have no enough knowledge about
the entanglement in the multipartite systems. Alternatively, the
numerical simulations should be able to be used for revealing and
analysis the behaviors of $E^{\rm ghz}(i,j)$ and $E^{\rm W}(i,j)$.
Nevertheless, we will try, in theoretical, to prove strictly that
our GHZ-class and W-class entanglement measure norms are invariant
or non-increasing under LOCC.

This study is on progressing.

\end{document}